\documentclass{article} 

\usepackage{amssymb,amsmath,amsfonts}
\usepackage{multirow}
\usepackage{multicol}
\usepackage{pifont}
\usepackage[utf8]{inputenc}
\usepackage[T1]{fontenc}
\usepackage{lmodern}
\usepackage{hyperref}

\newtheorem{theorem}{Theorem}

\title{
 Electoral Systems Used around the World%
 \footnote{This is a personally archived version of a chapter by the same title~\cite{elecsys16} contributed to the book ``Real-World Electronic Voting: Design, Analysis and Deployment'', Feng Hao and Peter Y.\ A.\ Ryan (editors), Series in Security, Privacy and Trust, CRC Press, 2016.}
}
\author{
 Siamak F.\ Shahandashti\\
 School of Computing Science, Newcastle University, UK
}

\begin{document}
\maketitle 

\begin{abstract}
We give an overview of the diverse electoral systems used in local, national, or super-national elections around the world. 
We discuss existing methods for selecting single and multiple winners and give real-world examples for some more elaborate systems. 
Eventually, we elaborate on some of the better known strengths and weaknesses of various methods from both the theoretical and practical points of view. 
\end{abstract}


\section{Introduction}
\label{sec:intro-elec-sys}

An \emph{electoral system}, or simply a \emph{voting method}, defines the rules by which the choices or preferences of voters are collected, tallied, aggregated, and collectively interpreted to obtain the results of an election~\cite{gallagher2005politics,robinson2010mathematical}. 

There are many electoral systems. 
A voter may be allowed to vote for one or multiple candidates, one or multiple predefined lists of candidates, or state their preference among candidates or predefined lists of candidates. 
Accordingly, tallying may involve a simple count of the number of votes for each candidate or list, or a relatively more complex procedure of multiple rounds of counting and transferring ballots between candidates or lists. 
Eventually, the outcome of the tallying and aggregation procedures is interpreted to determine which candidate wins which seat. 

Designing end-to-end verifiable e-voting schemes is challenging. 
Indeed, most such schemes are initially designed to support relatively unsophisticated voting methods in which ballot structure and tallying rules are straightforward. 
However, extending such a scheme to support more complex voting methods may not be trivial. 
Issues such as efficiently encoding preferential ballots with a large number of candidates and preserving voter privacy when transferring ballots during multiple rounds of counting can introduce considerable design challenges. 
Such challenges are evidenced for instance by the compromises made in the design of the state-of-the-art vVote system used for recent Victorian elections~\cite{vVoteTISSEC}. 
There have been a few works attempting to address these challenges (see, e.g.,~\cite{xia2010versatile} and the references within), nevertheless achieving practical end-to-end verifiable schemes supporting complex voting methods remains an area of research with many open questions. 
A good understanding of how different voting methods work is a prerequisite for tackling such open questions. 
In this chapter we aim to provide an introduction to the diverse voting methods used around the world. 

Mathematically, an electoral system can be seen as a \emph{function} that takes as input the choices or preferences of the voters and produces as output the results of the election. 
Voting theory, and more broadly \emph{social choice theory}, provides a formal framework for the study of different electoral systems, and in general social choice functions. 
A social choice function in this framework is a function that takes as input a set of individual orderings of a set of alternatives and produces a social ordering of the alternatives. 
This formalization was first put forth by Arrow~\cite{arrow51social}, a pioneer of modern voting theory. 

In practice however, there is much more to an election than just the electoral system, and these other issues are equally (if not more) important as the choice of the electoral system in ensuring fair and free elections and establishing public trust. 
Among these issues are (pre-election) voter registration, observer missions during the election, and post-election audits. 
From a legal point of view, the electoral system is only one part of the much wider electoral laws and regulations which govern the rules and procedures involved in calling, running, and finalizing an election from the start to the end. 
These rules and procedures include those of voter eligibility, candidate nomination, party campaigning, election administration, and announcement of results. 
In this chapter however, we mainly focus on electoral systems. 

Electoral systems can be categorized in multiple different ways. 
Two common criteria for categorization are whether the system is designed to produce one winner or multiple winners, and whether the system is designed to produce results that are roughly proportional to the vote share of each party or the system is based on the ``winner takes all'' approach. 
In the remainder of this chapter however we have chosen not to be bound to such categorizations. 
Instead, we follow the ideas underlying different electoral systems and work our way from the more immediate design ideas to the more elaborate ones. 

\section{Some Solutions to Electing A Single Winner}
\label{sec:sinwin}

Perhaps the most natural solution to elect a single winner is to elect the candidate with the most votes. 
This idea is the basis of the so-called first-past-the-post electoral system. 

In a \textbf{first-past-the-post (FPTP)} system, each voter can vote for one candidate and the single candidate with the highest number of votes wins. 
The winner might achieve an absolute majority of votes (i.e., more than a half), or merely a plurality of votes (i.e., most votes relatively). 
The system is also known as \textbf{single-member plurality (SMP)} or \textbf{simple plurality}. 
In the case of a race with only two candidates such a system is also called a \textbf{simple majority} system. 

First-past-the-post is used, among other places, in USA presidential elections (48 states)~\cite{bowler2005united}, UK lower house elections~\cite{mitchell2005united}, Canada~\cite{massicotte2005canada}, India~\cite{heath2005india}, and Malaysia~\cite{hai2002electoral}. 

There are variants of the first-past-the-post system that require the winning candidate to achieve a \emph{quota}, i.e., a threshold of votes, which is higher than the natural quota. 
For instance, in a two-candidate election, the winning candidate might be required to receive a quota which is greater than half of the votes: in the United States upper house, a so-called filibuster preventing legislation may be stopped only if the legislation receives three-fifth of the votes~\cite{fisk1997filibuster}. 
These systems are sometimes called \textbf{quota} systems, and in the case of a two-candidate election a \textbf{super-majority} system. 

Note that in the first-past-the-post system, each voter is restricted to vote for only one candidate. 
If this restriction is lifted, the resulting system is called approval voting. 

In an \textbf{approval voting} system, each voter may vote for (i.e., approve of) any number of candidates and the single candidate with the highest number of votes (i.e., approvals) wins. 

Approval voting is used among other places by the Mathematical Association of America~\cite{maa}, the Institute for Operations Research and the Management Sciences~\cite{informs}, and the American Statistical Association~\cite{amstat}. 

Although first-past-the-post provides a simple solution to elect a single winner, it does not guarantee an absolute majority if there are more than two candidates. 
One way to make sure that the winner receives an absolute majority is to choose the two candidates with the most votes for a second round of voting. 

In a \textbf{two-round system (TRS)}, each voter votes for one candidate. 
If a candidate receives more than half of the votes, they are declared the winner. 
Otherwise, the two candidates with the highest number of votes are chosen as the only candidates for a second round of voting, and the rest of the candidates are eliminated. 
In the second round, each voter can vote for one of the two remaining candidates, and the candidate with the highest number of votes wins.
The system is sometimes abbreviated as \textbf{2RS} and is also known as \textbf{run-off voting} and \textbf{double-ballot}. 

The two-round system is used in many countries to elect members of the parliament and directly-elected presidents, e.g., in both presidential elections and lower house elections in France~\cite{elgie2005france}. 

There are other variations of TRS in which all candidates receiving a certain quota become eligible for the second round, or a candidate can be declared a winner in the first round if they meet certain conditions, e.g., achieve a certain quota and have a certain lead over the second candidate. 

To avoid the cost of a second round of voting, an idea is to ask voters for their preferences between the candidates on the ballot. 

In the \textbf{contingent vote} system, voters rank the candidates in order of preference. 
The ballots are then distributed between the candidates based on their first preference votes. 
If a candidate receives more than half of the ballots (i.e., the first preference votes), they are declared the winner. 
Otherwise, the two candidates with the highest number of first preference votes are chosen as the only candidates for a second round of counting, and the rest of the candidates are eliminated. 
In the second round of counting, the ballots stating an eliminated candidate as the first preference are re-distributed (or transferred) to one of the two remaining candidates based on which candidate is ranked above the other. 
Eventually, the candidate with the highest number of votes is declared the winner. 

A variant of the contingent vote where the voters are restricted to express only their top two preferences is used to elect the directly elected mayors in England, including the Mayor of London~\cite{uk2015voting}. 
Another variant where the voters are restricted to express only their top three preferences is used in the Sri Lankan presidential elections~\cite[p.~135]{reilly2005electoral}. 
Note that these variants do not guarantee an absolute majority for the winner. 

An alternative to ensure an absolute majority for the winner is to carry out multiple rounds of voting and in each round only eliminate the candidate with the lowest number of votes. 

In the \textbf{exhaustive ballot} system, the voter may vote for one candidate of their choice in each round of voting. 
If a candidate receives an absolute majority of the votes, they are declared the winner. 
Otherwise, the candidate with the lowest number of votes is eliminated and the next round of voting is carried out between the remaining candidates. 
These steps are repeated until a candidate receives an absolute majority. 

The exhaustive ballot system is used among other places to elect the members of the Swiss Federal Council~\cite{SwissFedCouncil}, the President of the European Parliament~\cite{EuParlPres}, the speakers of the Canadian House of Commons~\cite{CanadaCommonsSpeaker}, the British House of Commons~\cite{UKCommonsSpeaker}, and the Scottish Parliament~\cite{ScotParlSpeaker}, the host city of the Olympic Games, and the host of the FIFA World Cup. 

To avoid multiple rounds of voting, the voters can be asked to state their preferences on the ballots. 
This is the basis for the following system. 

In the \textbf{instant run-off voting (IRV)} system, the voters rank the candidates in order of preference. 
The ballots are then distributed between the candidates based on their first preference votes. 
If a candidate receives more than half of the ballots (i.e., the first preference votes), they are declared the winner. 
Otherwise, the candidate with the lowest number of allocated ballots is eliminated and their allocated ballots are redistributed (or transferred) to the next ranked candidate on each ballot who is not yet eliminated. 
These steps are repeated until a candidate is allocated an absolute majority of the ballots and is declared the winner. 
The system is also known as the \textbf{alternative vote (AV)}. 

The instant run-off electoral system is used among other places in the Australian lower house elections~\cite{farrell2005australia}, and the Irish presidential elections~\cite{roi-president}. 

Partial ranking of the candidates might be allowed. 
In this case, all the candidates ranked on a ballot might get eliminated before the final round. 
Such ballots are called \emph{exhausted} ballots. 
The system guarantees an absolute majority only among the ballots that are neither spoiled nor exhausted by the last round of counting. 
On the other hand, voters might be asked to submit a full ranking of all the candidates on the ballot so as to minimize exhausted and hence ``wasted'' ballots. 
However, in practice this usually leads to an increase in the number of invalid votes. 

\begin{table}
  \caption{An example of instant run-off voting (IRV): The 2014 mayoral election results in Derwent Valley council, Tasmania, Australia}
  \label{tbl:der}
  \centering 
  \fbox{
  \begin{tabular}{rrrrr|rcl}
    \multicolumn{5}{c|}{Candidates} & & & \\ 
    PBe & PBi & MEv & CLe & FPe & Exhausted & Majority & Remark \\ \hline 
     870 &  333 & 1632 &  423 &  620 &    0~~~~ & 1940 & Count 1\\ 
     +73 & -333 &  +86 &  +60 &  +62 &  +52~~~~ &      & PBi excluded \\ \hline 
     943 &    0 & 1718 &  483 &  682 &   52~~~~ & 1914 & Count 2\\ 
    +154 &      & +147 & -483 & +135 &  +47~~~~ &      & CLe excluded \\ \hline 
    1097 &      & 1865 &    0 &  817 &   99~~~~ & 1890 & Count 3 \\ 
    +386 &      & +307 &      & -817 & +124~~~~ &      & FPe excluded \\ \hline 
    1483 &      & \underline{2172} &      &    0 &  223~~~~ & 1828 & Count 4 \\ 
         &      &      &      &      &      &      & MEv elected \\ 
  \end{tabular}
  }
\end{table}

As an example of IRV, consider the results shown in Table~\ref{tbl:der} for the election of mayor in Derwent Valley council from the 2014 Tasmanian local government elections~\cite{derres}. 
There were a total of 3878 valid ballots, which means the initial quota for absolute majority was $\lfloor 3878/2 \rfloor +1=1940$, where $\lfloor \cdot \rfloor$ denote the floor function. 
The first five columns show the progressive total ballots for the five candidates. 
As seen in the table, in the first count no candidate achieves absolute majority, and hence the candidate with the lowest number of votes, PBi, is eliminated. 
PBi's 333 ballots are examined and transferred to their respective second preferences: in this case, 73 to PBe, 86 to MEv, 60 to CLe, and 62 to FPe. 
Fifty two ballots do not have a second preference stated, and hence are exhausted. 
This means that in the next round the quota for absolute majority is reduced to 1914. 
No candidate achieves majority in the second and third rounds of voting and further two candidates are eliminated and their ballots transferred. 
In the final round, MEv has 2172 ballots which is above the absolute majority quota of 1828 and hence MEv is elected. 

The IRV method discussed above is the single-winner version of an electoral system known as the single transferable vote (STV) which we will discuss later in this chapter. 
These methods were proposed by Thomas Hare~\cite{hare1857machinery}, and hence are sometimes collectively known simply as the Hare system. 

While Hare's method eliminates the candidate with the lowest first-preference votes in each round, a variant called \textbf{Coombs' method}~\cite{coombs1964theory} eliminates the candidate with the highest last-preference votes in each round. 
In other words, in each round Hare excludes the least liked candidate, whereas Coombs excludes the most disliked candidate.

\section{Some Solutions to Electing Multiple Winners}
\label{sec:mulwin}

To elect multiple winners, one could of course simply extend the first-past-the-post system and elect multiple candidates with the highest number of votes. 
Let us assume the desired number of winners (or seats) is $n$. 

In a \textbf{block-vote (BV)} system, a voter votes for up to $n$ candidates. 
The candidates are then ordered based on the number of votes they have received and the first $n$ candidates are declared winners. 
The system is also known as \textbf{plurality-at-large voting} and \textbf{multiple non-transferable vote (MNTV)}. 

The system is used among other places in elections in Lebanon~\cite{salloukh2006limits}. 

The \textbf{single non-transferable vote (SNTV)} system can be seen as a block-vote system in which the voters are restricted to vote for only one candidate. 
This system is used among other places in the Japanese upper house elections~\cite{cox2000electoral}. 

A block-vote system in which the voters can vote for more than one but fewer than $n$ candidates is known as \textbf{limited vote (LV)}. 
The Spanish upper house elections use this system, in which the voters may vote for up to three candidates whereas four winners are elected~\cite{hopkin2005spain}. 

Another variant of the block-vote, sometimes called the \textbf{party block-vote (PBV)}, requires voters to vote for a party (or in general a predetermined list of candidates) instead of voting directly for candidates. 
After the count, the party with the highest number of votes is allocated all the $n$ seats. 
This variant can be thought of a first-past-the-post election between parties. 
It is used among other places in Cameroon~\cite[Annex~A]{reilly2005electoral} and Singapore~\cite[Annex~A]{reilly2005electoral}. 

The party block-vote system, like many other systems based on the ``winner takes all'' paradigm, may produce results that are significantly skewed towards one or more popular parties. 
The underlying idea of the so-called \textbf{proportional representation (PR)} electoral systems is to ensure that the number of elected candidates from each party (or coalition of parties) is to some extent proportional to their respective share of the votes. 

In the \textbf{list voting} or more specifically \textbf{party-list PR} system, each party presents a list of candidates and seats are allocated to each party in proportion to the number of votes the party receives. 

In what is known as the \textbf{closed-list} variant, the voters vote for a list, and after the number of seats allocated to each party is determined, that number of candidates on top of the party list are elected. 
Hence, the order in which candidates get elected from each list is pre-determined merely by the party and the voters do not get to choose it. 
The closed-list system is used among other places in national parliamentary elections in Argentina~\cite[Annex~A]{reilly2005electoral}, Portugal~\cite{costa2012political}, Spain~\cite{hopkin2005spain}, and South Africa~\cite{gouws2005south}. 
The system is also used in the European parliament elections in many countries including Germany, France, United Kingdom (excluding Northern Ireland), and Spain~\cite[Part~5]{strauch2011electoral}. 

On the other hand, in the \textbf{open-list} variant, the voters vote for candidates, and the number of votes each candidate receives influences the order in which candidates are chosen from a party list at the end of the election. 
Since voter preferences can influence the order of the elected candidates, such systems are also known as \textbf{preferential} list voting. 
There are multiple different deployments of this variant which give the voter varying amounts of influence. 
Hence, some scholars suggest using the term ``open list'' exclusively for the systems in which the order of elected candidates is solely determined by voter preferences, and refer to the systems in which the order of elected candidates is determined by a combination of party list orders and voter preferences as \textbf{flexible} list voting (see, e.g.,~\cite{shugart2001mixed,gallagher2005politics}). 
Open-list voting is used widely around the world including in the Brazilian~\cite{nicolau2004brazil}, Dutch~\cite{andeweg2005netherlands}, Czech~\cite{kopecky2004czech}, and Swedish~\cite{sarlvik2007sweden} lower house elections. 
The system is also used in the European parliament elections in many countries including Italy, Poland, and the Netherlands~\cite[Part~5]{strauch2011electoral}. 
The open-list systems used in Luxembourg and Switzerland parliamentary elections are unique in that they allow for \emph{panachage}, i.e., voters are allowed to split their preferences between multiple parties~\cite{kolk2007local,lutz2010first}. 

A two-round variant of the closed-list system is in use in French regional elections~\cite{lutz2010first}. 
Any party with at least a predetermined threshold of the votes may contest the second round. 
In the second round, the seats are allocated to the parties proportionally to their shares of the votes. 

There are various methods for seat allocation based on each party's share of the votes. 
The two most common categories are the highest average and the largest remainder methods. 

In the \textbf{highest-average (HA)} methods, the number of votes for each party is successively divided by a set of \emph{divisors}, resulting in a series of quotients called \emph{averages}. 
Eventually, $n$ of the top values among the averages of all parties are determined and the number of averages selected for each party gives their share of the final $n$ seats. 

\begin{table}
  \caption{An example of the d'Hondt method of proportional representation voting: 2011 election results in the Cz\k{e}stochowa constituency, Poland}
  \label{tbl:cze}
  \centering 
  \fbox{
  \begin{tabular}{lrrrrc} 
	Party & vote & vote/2 & vote/3 & vote/4 & seats \\ 
	\hline 
	PO	& \underline{34.97}	& \underline{17.49}	& \underline{11.66}	& 8.74 & 3 \\ 
	PiS	& \underline{27.36}	& \underline{13.68}	& 9.12	& 6.84 & 2 \\ 
	RP	& \underline{13.39}	& 6.70	& 4.46	& 3.35 & 1 \\ 
	SLD	& \underline{10.49}	& 5.25	& 3.50	& 2.62 & 1 \\ 
	PSL	& 8.77	& 4.39	& 2.92	& 2.19 & 0 \\ 
	PJN	& 2.14	& 1.07	& 0.71	& 0.54 & 0 \\ 
	NP	& 2.06	& 1.03	& 0.69	& 0.52 & 0 \\ 
	PPP	& 0.84	& 0.42	& 0.28	& 0.21 & 0 \\ 
  \end{tabular}
  } 
\end{table}

One of the most widely used highest-average methods is the \textbf{d'Hondt} method in which the divisors are $(1,2,3,4,\ldots)$. 
The method is used among many other places in the Polish lower house elections~\cite{moser1999electoral}. 
Table~\ref{tbl:cze} shows the results for the lower house constituency of Cz\k{e}stochowa in the 2011 Polish parliamentary elections according to the Polish national electoral commission~\cite{czeres}. 
The constituency has seven seats. 
The first two columns show the parties with their (rounded) percentage of valid votes. 
The third, fourth, and fifth columns show the votes for each party divided by the divisors 2, 3, and 4, respectively. 
The seven highest averages in the table, shown underlined, determine the number of seats allocated to each party. 
For instance, since the PiS party has 2 of the highest 7 averages, it wins 2 of the 7 seats. 
The idea here is that any change in the allocated number of seats would put a party in disadvantage in terms of average number of votes per seat. 
For example, PO's 3 seats means they have a seat on average for every 11.66\% of votes, whereas if PO's third seat were allocated to PSL instead, it would mean that PSL would get a seat on average for every 8.77\% of the votes. 


The \textbf{Sainte-Lagu\"{e}} method is similar to the d'Hondt method, but uses the divisors $(1,3,5,7,\ldots)$ instead. 
Other highest average methods also follow the same principle, but utilize different divisors. 
Among these are the \textbf{modified Sainte-Lagu\"{e}} method with divisors $(1.4,3,5,7,\ldots)$, the \textbf{Imperiali} method with divisors $(2,3,4,5,\ldots)$, and the \textbf{Danish} method with divisors $(1,4,7,10,\ldots)$. 

In the \textbf{largest remainder (LR)} methods, first a \emph{quota} is calculated, representing the number of votes required for a seat. 
Then the number of votes for each party is divided by the quota to obtain a quotient consisting of an integer and a fractional part. 
The fractional part is called a \emph{remainder}. 
Each party is allocated a number of initial seats equal to the integer part of their quotient. 
This will amount to a total of $n_\mathrm{i}$ initial seats. 
The remaining $n-n_\mathrm{i}$ seats are distributed between the $n-n_\mathrm{i}$ parties with the largest remainders, giving each such party an extra seat. 

The \textbf{Hare} quota and the \textbf{Droop} quota are two widely used quotas in LR systems. 
The Hare quota is calculated by dividing the total number of (valid) votes to the number of seats. 
The Droop quota is calculated by dividing the total number of (valid) votes to the number of seats plus one, and then adding 1 to the result. 
Fractions are usually disregarded in calculating quotas. 
In other words, we have: 
\[ 
 \text{Hare quota}=\left\lfloor \frac{\text{no. of votes}}{\text{no. of seats}} \right\rfloor
 \quad\text{and}\quad 
 \text{Droop quota}=1+ \left\lfloor \frac{\text{no. of votes}}{1+\text{no. of seats}} \right\rfloor 
 \ . 
\] 
Other quotas that are used include the  \textbf{Hagenbach-Bischoff} quota which is one less than the Droop quota, and the \textbf{Imperiali} quota which is calculated by dividing the number of votes into the number of seats plus two. 

The Droop quota is used in the national and provincial elections in South Africa~\cite{gouws2005south}. 
Table~\ref{tbl:gau} shows the Gauteng Provincial Legislature results in the 2014 South African National and Provincial Elections~\cite{gaures}. 
The table only shows the first eight parties. 
The first two columns show the parties and their respective number of votes. 
There are 73 seats to be allocated. 
The total number of valid votes is 4,382,163. 
The Droop quota hence is calculated as $\lfloor 4,382,163/(73+1) \rfloor +1 = 59,219$. 
Dividing the votes for each party by the quota gives the quotient, the number of initial seats (the integer part of the quotient), and the remainder (the fractional part of the quotient). 
The total number of initial seats is 68 which leaves 5 extra seats to be allocated to the 5 parties with the largest remainders, shown underlined in the table. 

\begin{table}
  \caption{An example of the largest remainder method of proportional representation using the Droop quota: 2011 Gauteng Provincial Legislature election results, South Africa}
  \label{tbl:gau}
  \centering 
  \fbox{
  \begin{tabular}{lrrrcrr}
	Party &	votes & quotient & initial & remainder & extra & total \\ 
	      &	      &          & ~seats~ &           & seats & seats \\ 
	\hline 
	ANC &	2,348,564 &		39.66 &	39~~ &	\underline{0.66} &	1~~~ &	40~~ \\ 
	DA &	1,349,001 &		22.78 &	22~~ &	\underline{0.78} &	1~~~ &	23~~ \\ 
	EFF &	  451,318 &	 	 7.62 &	 7~~ &	\underline{0.62} &	1~~~ &	 8~~ \\ 
	VF+ &	   52,436 & 	 0.89 &	 0~~ &	\underline{0.89} &	1~~~ &	 1~~ \\ 
	IFP &	   34,240 & 	 0.58 &	 0~~ &	\underline{0.58} &	1~~~ &	 1~~ \\ 
	ACDP &	   27,196 & 	 0.46 &	 0~~ &	           0.46  &	0~~~ &	 0~~ \\ 
	COPE &	   21,652 & 	 0.37 &	 0~~ &	           0.37  &	0~~~ &	 0~~ \\ 
	NFP &	   20,733 & 	 0.35 &	 0~~ &	           0.35  &	0~~~ &	 0~~ \\ 
  \end{tabular}
  }
\end{table}

The South African system is an example of a PR system without a \emph{threshold}. 
However, most PR systems require a threshold to be achieved for the party to be eligible for any seats. 
The lower the threshold is, the more proportional the results will be. 

Some argue that in many of the systems discussed so far, especially if the number of seats is relatively low, there is a potential for many votes to be so-called ``wasted''. 
For example, in Table~\ref{tbl:cze}, votes for the last four parties, although counting for more than 10\% of the votes, do not count toward electing any candidate and are arguably wasted. 
The single transferable vote (STV) system, which can be seen as a generalization of the instant run-off (IRV) to elect multiple winners aims to minimize votes being wasted by asking voters to declare their preferences. 
This way, if a preferred candidate does not receive enough support to be elected, the vote is transferred to the next preferred candidate and finally counts towards electing one of the candidates on the voter's list. 
STV was first proposed in the 1850s by Thomas Hare~\cite{hare1857machinery}. 

In the \textbf{single transferable vote (STV)} system, the voters indicate their preferences between the candidates by ranking them on the ballot. 
In each round of counting if a candidate achieves a certain \emph{quota}, he or she is elected. 
Otherwise, the candidate with the lowest number of votes is eliminated from the race. 
Then either the elected candidate's \emph{surplus} votes or all of the eliminated candidate's votes are transferred to the next candidate appearing on the preference list who is neither already elected nor already eliminated. 
The process continues until either all seats are allocated or the number of candidates remaining in the race is reduced to the number of remaining available seats. 

STV is used in parliamentary elections in Ireland~\cite{gallagher2005ireland} and the upper house elections in Australia at the national level~\cite{farrell2005australia}, and in the Scottish local council elections~\cite{bennie2006transition} and Tasmanian lower house elections~\cite{farrell1996designing} at the subnational level. 
The system is also used in the European parliament elections in Ireland, Northern Ireland, and Malta~\cite[Part~5]{strauch2011electoral}. 

The quota normally used with STV is the Droop quota. 
Transferring ballots for the eliminated candidates is similar to that of IRV. 
However, in case a candidate achieves higher votes than the quota, their ballots above the quota are called a surplus and may be transferred. 
One may think of this process as transferring a portion of the elected candidate's ballots that are not needed for them to be elected. 
Hence, all the transferable ballots are examined, and the share of each next preference from the surplus votes is determined. 
This usually results in fractional ballot transfers between the candidates. 
The rules governing when and how exactly the surplus transfers should be carried out are different between elections in different countries. 

Determining STV election winners can be complex and often consists of tens of rounds of counting. 
Here however we consider a less complex example. 
Table~\ref{tbl:abo} shows the results of the 2009 Aboriginal Land Council of Tasmania elections in the South Region~\cite{abores}. 
The total number of votes is 71, and two candidates are to be chosen. 
Hence, the initial Droop quota is $\lfloor 71/3 \rfloor +1 = 24$. 
As the table shows, in the first round, candidate S's first preference votes are more than enough to get him or her elected. 
Thus, S is declared elected in the first round. 
However, since S only needs 24 votes to get elected, S's surplus votes, 9 votes in this case, are transferred to his or her corresponding second preferences. 
To do this fairly, all the 33 votes are examined. 
In this case, 6 of S's ballots list K as the second preference, 15 list M, and 12 list N. 
That is, $6/33$ of any transferring ballot should go to K, $15/33$ to M, and $12/33$ to N. 
Now that 9 ballots are transferring, $9(6/33) \approx 1.63$ ballots go to K, $9(15/33) \approx 4.09$ to M, and $9(12/33) \approx 3.27$ to N. 
The totals in the second round do not push any candidate above the quota, hence the candidate with the least votes, K, is eliminated and K's 7 votes are distributed, in this case, 4 to M and 3 to N. 
This gives N enough votes to be declared the second winner. 

\begin{table}
  \caption{An example of single transferable vote (STV) using the Droop quota: 2009 Aboriginal Land Council of Tasmania (South) election results, Australia}
  \label{tbl:abo}
  \centering 
  \fbox{
  \begin{tabular}{r@{}lr@{}lr@{}lr@{}l|ccl}
    \multicolumn{8}{c|}{Candidates} & & & \\ 
    \multicolumn{2}{c}{K} & \multicolumn{2}{c}{M} & \multicolumn{2}{c}{N} & \multicolumn{2}{c|}{S} 
                          & Exhausted & Quota & Remark \\ \hline 
     7 &     &  13 &     & 18 &     & \underline{33} & & 0 & 24 & Count 1 \\ 
    +1 & .63 &  +4 & .09 & +3 & .27 & -9 & &   &    & S elected \\ \hline 
     8 & .63 &  17 & .09 & 21 & .27 & 24 & & 0 & 24 & Count 2 \\ 
    -7 &     &  +4 &     & +3 &     &    & &   &    & K excluded \\ \hline 
     1 & .63 &  21 & .09 & \underline{24} & .27 & 24 & & 0 & 24 & Count 3 \\ 
       &     &     &     &    &     &    & &   &    & N elected \\   
  \end{tabular}
  }
\end{table}

The above example was a rather straightforward case of determining STV winners. 
However, note that in many cases for instance if there are multiple winners in any round, or if there are exhausted ballots and hence the quota changes, there could be different methods for how and when to transfer votes. 
Although the difference between such different methods might seem insignificant, they may lead to different outcomes in the election. 
The transfer rules are usually agreed on and published in detail before the election, and as mentioned before, they vary considerably between different jurisdictions. 

\section{Blending Systems Together}
\label{sec:mixsys}

Elections with single-member districts are praised for clearly tying a representative to a constituency and hence fostering a higher degree of accountability for elected representatives. 
On the other hand, elections with multi-member districts using proportional representation (PR) systems such as party-list are designed to produce results in which the number of seats each party wins is to a great extent proportional to the party's share of popular vote. 
To combine the positive aspects of these two types of systems, many jurisdictions run two systems alongside each other. 

In a \textbf{Mixed Member Proportional (MMP)} system, one voting method is used for electing individual representatives for each constituency, and besides this first method, a second PR method is used to compensate for any disproportionality produced by the constituency results. 
In some MMP systems, the voter is able to vote in each method separately. 
In other systems however, the voter votes for the constituency representative only, and the party vote is calculated by aggregating the candidate votes in all of the constituencies in a larger PR district.
There may be a single national PR district or several subnational ones. 


The MMP system is used among other places in parliamentary elections in Germany~\cite{saalfeld2005germany}, Hungary~\cite{benoit2005hungary}, Mexico~\cite{diaz2004mexico}, and New Zealand~\cite{vowles2005new}, which use combinations of first-past-the-post and list-PR. 

In a \textbf{two-tier} system, two parallel and independent methods are used: one voting method is used for electing individual representatives for each constituency, and a PR method is used to elect members proportional to party vote shares independently of how many seats the parties win at constituency level. 
The PR method districts are larger than the constituencies, usually several subnational districts or a single national district. 
Two-tier systems are also known simply as \textbf{parallel} systems. 

The two-tier system is used in parliamentary elections among other places in South Korea~\cite{rich2013evaluating}, Japan~\cite{reed2005japan}, and Thailand~\cite{hicken2005thailand}, which use first-past-the-post alongside list-PR, and in Lithuania~\cite{moser1999electoral}, which uses the two-round system alongside list-PR.

\section{Other Solutions}
\label{sec:miscvotsys}

In this section, we review some of the other systems that are less widely used in national and subnational elections. 

In the \textbf{Borda count}, each voter ranks the candidates on the ballot. 
The candidates each get a number of points based on their rank, according to a point allocation scheme which is decreasing with respect to rank. 
For instance, if there are $k$ candidates on the ballot, the $i$-th ranked candidate is allocated $k-i$ points, i.e., $k-1,k-2,\ldots,0$ points respectively for candidates in the order of preference. 
The points each candidate receives in all ballots are summed up and the candidate with the highest sum of points is declared the winner. 

This system is used in a few political elections around the world including Nauru~\cite{reilly2002social} and Kiribati~\cite{reilly2002social}, and other places such as the Eurovision Song Contest~\cite{eurovision2015rules}. 
In Slovenia, the Borda count, which is used to elect the representatives for the Hungarian and Italian-speaking ethnic minorities, allocates $k+1-i$ points to the $i$-th ranked candidate, i.e., $k,k-1,\ldots,1$ points respectively for candidates in the order of preference. 
In parliamentary elections in Nauru, the $i$-th ranked candidate is allocated $1/i$ points, i.e., $1,\frac{1}{2},\ldots,\frac{1}{k}$ points, respectively, for candidates in the order of preference. 


In the \textbf{cumulative voting} system, each voter has a fixed number of points to share between a number of candidates, and the single or multiple candidates receiving the highest total points are declared winners. 

Cumulative voting is used among other places in Norfolk Island Legislative Assembly elections where each voter gets nine votes to share between the candidates with the restriction that no more than two votes can be given to any single candidate~\cite{bennett2007australian}. 
Besides, the system is used in some local elections in the United States (see, e.g.,~\cite{CumulUSA,CumulTexas}), and also in board elections in corporate governance (see, e.g.,~\cite[p.~270]{braendle2006shareholder}), where typically each shareholder is given a number of votes proportional to their share. 

In a \textbf{range voting} system, the voter rates the candidates on the ballot, i.e., gives each a score, and the candidate with the highest sum of scores is declared the winner. 
Approval voting can be seen as an instance of range voting in which only binary scores, i.e., approve or disapprove, are allowed. 
A variant called \textbf{majority judgement} calculates the winner based on the median score for each candidate. 

Range voting is used in scoring some sports competitions such as figure skating~\cite{FigureSkating} and gymnastics~\cite{gymnastics} where a truncated mean of the scores from multiple judges determines the final results. 
It is also used in web-based scoring and recommendation systems such as the Internet Movie Database (IMDb) where a weighted mean of the individual scores determines the final scores~\cite{imdb}. 

In \textbf{Condorcet methods}, the voter usually ranks the candidates, and Condorcet winner is the candidate, if any, which is pair-wise preferred to all other candidates by the majority of voters. 
The Condorcet winner is not guaranteed to exist. 
Any method that elects the Condorcet winner, if any, is generally known as a Condorcet method. 
A Condorcet method for $n$ candidates can be thought of as running $\frac{1}{2}n(n-1)$ simple majority elections between all possible pairs of candidates, and finding if there is a candidate that beats all others in their corresponding head-to-head election. 

There are various methods to calculate the Condorcet winner if any, and otherwise produce a plausible replacement winner. 
For instance, in the method known as \textbf{Smith/IRV}, the counting produces a so-called \textit{Smith set}, defined as the smallest non-empty set of candidates such that every candidate in the set defeats every candidate outside the set in a pair-wise election. 
The Condorcet winner is guaranteed to be in the Smith set. 
Hence, if the Smith set includes only one candidate, that candidate is declared the Condorcet winner. 
If the Condorcet winner does not exist, then the IRV method is used to elect a winner between the candidates in the Smith set. 

In the system known as \textbf{Black's method} if the Condorcet winner exists, they are declared the winner, and otherwise the Borda count is used to calculate the winner. 

Another Condorcet method known as the \textbf{Schulze method}~\cite{schulze2011new} involves finding preference paths between candidates and comparing them based on the so-called ``strength'' of the paths. 
The method outputs a complete ordering of the candidate and hence can be used to elect multiple candidates. 

Condorcet methods, and specifically the Schulze method, are fairly popular within the free software and free culture communities, and for instance are used in the internal elections of the several national Pirate Parties~\cite[p.~213]{rothe2015economics}, the Debian project~\cite{debian}, Ubuntu~\cite{ubuntu}, KDE~\cite{kde}, and the Free Software Foundation Europe~\cite{fsfe}.

\section{Which Systems Are Good?}
\label{sec:syscomp}

Every one of us might have already had a favourite electoral system before reading this chapter, or might  have set our mind on one while reading the chapter. 
We might think that our favourite system is obviously superior to the others we know of and have our reasons supporting our argument. 
However, social choice theorists on the one hand and electoral assistance experts on the other hand would be able to provide a variety of counter arguments pointing towards the weaknesses of our favourite system compared to other systems. 
In this section we aim to go through some of the better known comparative strengths and weaknesses of the electoral systems we have discussed, from both the theoretical and practical points of view.

\subsection{A Theorist's Point of View}
\label{sec:syscompth}

Social choice theory provides a variety of results on the merits of different electoral systems. 
Some of these results are naturally expected, while some utterly unexpected. 
Nonetheless, the results are interesting on both sides, either providing a solid theoretical foundation to build upon in the former case, or challenging our common understanding of such systems and compelling us to rethink and design better systems in the latter case. 

\subsubsection{Majority Rules}
\label{sec:majmay}

Let us first limit our attention to elections with only two candidates. 
Perhaps one of the expected, and yet illuminating early results in this case is May's theorem, which pretty much settles the question of which system is the best choice in elections with two candidates. 
To define a notion of a good system, let us start by defining the following criteria: 
\begin{itemize}
 \item 
 a system is called \emph{egalitarian}\footnote{This criterion is often called \emph{anonymity} in modern social choice theory. We use May's original term to avoid confusion with anonymity from the security viewpoint.} if it treats all voters equally; 
 \item 
 a system is called \emph{neutral} (with respect to candidates) if it treats all candidates equally; 
 \item 
 a system is called \emph{monotone} if the candidate who wins an election would still win if one or more voters change their vote in favour of the winning candidate and everyone else votes the same way;
 in other words, it is impossible for a winning candidate to become a losing candidate by gaining votes; and 
 \item 
 a system is called \emph{nearly decisive} if the only way a tie can occur is when the two candidates receive exactly the same number of votes. 
\end{itemize}
The above criteria seem quite natural to expect from a good electoral system. 
In fact, May has shown that the simple majority system is the only system that can satisfy all four criteria~\cite{may1952set}. 
\begin{theorem}[May's theorem]
In an election with two candidates, the only electoral system that is egalitarian, neutral, monotone, and nearly decisive is the simple majority method. 
\end{theorem}
May's theorem is definitive in that the simple majority system is the only system that could satisfy the above reasonable requirements. 
In fact, even if we do not care about the electoral system being decisive, an extension of May's theorem states that the only two-candidate electoral systems that are egalitarian, neutral, and monotone are the following ones: simple majority, super-majority, and a third nonsensical system which results in a tie regardless of the number of votes for the two candidates~\cite[p.~20]{robinson2010mathematical}. 
On the other hand, if we define a (strictly) decisive system to be one that always produces a winner (i.e., never ends in a tie), then it is not hard to see that the three properties of equality, neutrality, and decisiveness are inherently contradictory; 
that is, there is no electoral system for two candidates that is egalitarian, neutral, and decisive. 
This statement is true even when elections with more than two candidates are considered. 
This leads us to believe that (strict) decisiveness might be too strong a requirement to expect from an electoral system. 

\subsubsection{Bad News Begins}
\label{sec:impconiia}

Now consider elections with more than two candidates and a single winner. 
Equality and neutrality can still be defined similarly. 
Equality can be formalized by requiring that the outcome of the election stays the same if any two voters exchange their ballots. 
Similarly, neutrality can be formalized by requiring that if candidate A is replaced with candidate B on all ballots, and vice versa, i.e., candidate B is also replaced with candidate A on all ballots, then the same replacements are replicated in the outcome of the election. 

Formalizing monotonicity in the case of more than two candidates needs to be elaborated on to define a precise sense of the voters changing their votes \emph{in favour of} the winning candidate. 
In the case of only two candidates, it is clear that this means changing a vote for the losing candidate to a vote for the winning candidate. 
For an election with more than two candidates, let us consider the rather general case where voters rank the candidates on the ballots. 
We can now specify what is meant by changing a vote in favour of the winning candidate as changing the rank of the winning candidate on a ballot with the rank of a losing candidate which is ranked higher than the winning candidate, and vice versa. 

Let us now define more criteria to assess our electoral system against. 
All of these are criteria that we would naturally want a good system to satisfy. 
\begin{itemize}
 \item 
 a system satisfies the \emph{majority} criterion if whenever a candidate receives a majority of the first preferences, the system elects B as the winner; 
 \item 
 a system satisfies the \emph{Condorcet} criterion if it elects the Condorcet winner whenever such a winner exists; 
 \item 
 a system satisfies the \emph{Pareto} criterion (also called \emph{unanimity}) if whenever every voter prefers candidate A to candidate B, the system does not elect B as the winner; and 
 \item 
 a system satisfies the \emph{independence of irrelevant alternatives (IIA)} criterion if the following holds: consider an election in which A is elected the winner, and a second election in which all voters rank A above or below B the same way they have done in the first election, but may change their preferences of other candidates; the system must not choose B as the winner in the second election; 
 in other words, IIA requires that the electoral system's preference between any two candidates depends only on the individual voters' preferences between those two candidates. 
\end{itemize}
Note that if a candidate receives a majority of first preferences, the candidate beats all other candidates in head-to-head contests, and hence is the Condorcet winner. 
Thus, the Condorcet criterion is a stronger criterion than the majority criterion, i.e., the Condorcet criterion implies the majority criterion. 
In fact, the Condorcet and IIA criteria are incompatible as stated by the following theorem~\cite[p.~55]{robinson2010mathematical}. 
\begin{theorem}
There is no electoral system for an election with more than two candidates that satisfies both the Condorcet and the independence of irrelevant alternatives (IIA) criteria. 
\end{theorem}

The above theorem is one of several impossibility results in social choice theory. 
Each of these results shows the impossibility of electoral systems satisfying a set of criteria simultaneously. 
Such results can be seen as a contributing reason why the debate over the merits of different electoral systems is far from settled. 
A fundamental issue with distilling a social preference from a set of individual preferences which eventually is responsible for many such results is the following observation. 

The \textbf{Condorcet paradox} is the observation that majority preferences can be ``irrational'' (specifically, intransitive), even when individual preferences are ``rational'' (specifically, transitive). 

To see an example of this paradox, consider an election with three candidates A, B, and C. 
Assume we have three voters whose preferences are as follows. 
The first voter prefers A to B, and B to C, and since we are assuming rational voters, also A to C; 
or in shorthand $A \succ B \succ C$. 
The second voter's preferences are $B \succ C \succ A$, and the third voter's $C \succ A \succ B$. 
Now the majority of voters prefer A to B, B to C, and C to A. 
This means that although the individual preferences are transitive, the majority preference is intransitive. 

\subsubsection{Arrow's Impossibility Theorem}
\label{sec:imparrow}

A well-known impossibility result which has been described as ``the single most important result in the history of voting theory''~\cite{hodge2005mathematics} is Arrow's impossibility theorem. 
Arrow considers electoral systems that provide a full ranking of the candidates as outcome.
He defines the following criteria in addition to the ones we have discussed so far: 
\begin{itemize}
 \item
 a system satisfies the \emph{unrestricted domain} criterion (or universality, the term originally used by Arrow) if it does not place any restriction other than transitivity on how voters can rank the candidates; 
 \item
 a system satisfies the \emph{non-imposition} criterion (or citizen sovereignty, the term originally used by Arrow) if its outcome is not restricted (i.e., not imposed) in any way other than being transitive; in other words, every transitive outcome is possible in the election depending on individual orderings; and 
 \item
 a system satisfies the \emph{non-dictatorship} criterion if there is no single voter (i.e., a dictator) whose vote determines the outcome of the election regardless of how others vote. 
\end{itemize}
Note that non-dictatorship is a weaker criterion than equality, i.e., equality implies non-dictatorship. 

Arrow's impossibility theorem basically says that the only unrestricted-domain electoral systems which are monotone and independent of irrelevant alternatives are either imposed or dictatorial~\cite{arrow1950difficulty}. 
\begin{theorem}[Arrow's impossibility theorem]
There is no electoral system for an election with more than two candidates that satisfies the unrestricted domain, monotonicity, and independence of irrelevant alternatives (IIA) criteria and is neither imposed nor dictatorial. 
\end{theorem}

Arrow's impossibility theorem is pretty strong in ruling out the possibility of existence of any fair electoral system that satisfies three reasonable criteria that one may expect from a good system. 
It can be even stated in a stronger form since monotonicity, IIA, and non-imposition together imply the Pareto criterion. 
In its stronger form, the theorem basically says unrestricted domain, Pareto, and IIA properties are incompatible~\cite{hodge2005mathematics}. 
\begin{theorem}[Arrow's impossibility theorem (strong form)]
There is no electoral system for an election with more than two candidates that satisfies the unrestricted domain, Pareto, and independence of irrelevant alternatives (IIA) criteria and is not dictatorial. 
\end{theorem}

Although Arrow's impossibility theorem states that certain desirable criteria are incompatible with each other, what it does not say is that there are no reasonable systems around. 
The question of choosing the right system hence becomes that of the choices we make between the desirable criteria to achieve a compromise. 

One possible compromise would be to consider systems in which the voter's ranking of candidates is restricted in some way, and hence the system does not support an unrestricted domain. 
Of course this should be done in a way that neutrality is still kept intact. 
An example of such a system is the approval electoral system in which candidate rankings on the ballots are restricted to either approval or lack thereof. 
By compromising on the unrestricted domain criterion, approval voting is able to achieve monotonicity, Pareto, and IIA. 
Note that the Condorcet paradox is absent in the setting of approval voting since collective preference, as defined by comparing the number of approvals for each candidate, is transitive. 

When faced with a choice between Pareto and IIA, the more accepted view seems to support a compromise on IIA. 
IRV and Borda are both examples of systems which do not restrict voter's rankings of candidates in any way and at the same time achieve Pareto and provide some guarantees comparatively weaker than IIA.

\subsubsection{Gibbard--Satterthwaite Impossibility Theorem}
\label{sec:impgibsat}

Consider a single-winner election with three candidates A, B, and C using the Borda count. 
Assume A and B are the only main contenders with a realistic chance of winning. 
Consider a voter, Alice, whose preferences are as follows: 1st A, then B, and C last. 
If Alice reflects her preferences as they are on the ballot box, i.e., she puts $A \succ B \succ C$ on the ballot, it is said that she votes \emph{sincerely}. 
However, knowing that the realistic race is only between A and B, it would make sense for Alice to mark $A \succ C \succ B$ on her ballot to give her first preference a better chance of winning. 
This would be a case of so-called \emph{strategic} or \emph{tactical} voting in which considering contextual information the voter misrepresents her preferences on the ballot to favour a candidate over a relatively less preferred candidate. 

It is often argued that an electoral system should ideally ensure that, no matter the contextual circumstances, the best voting strategy for a voter always is voting sincerely, i.e., reflecting their actual preferences. 
However, a significant theoretical result known as the Gibbard--Satterthwaite theorem rules out the existence of such ideal electoral systems altogether under some natural conditions. 
In the following we briefly discuss this theorem. 

Let us for the moment limit our attention to single-winner systems only. 
A basic fairness criterion is to require that every candidate should be able to win. 
In the following the definition of this criterion is listed along with that of strategy-proofness. 
\begin{itemize}
 \item
 a system is said to have an \emph{unrestricted range} if its winner can be any candidate; and 
 \item 
 a system is said to be \emph{strategy-proof} (or non-manipulable) if there are circumstances under which strategic voting by a voter leads to a winner which is actually preferred by the voter to a candidate that will win if the voter votes sincerely. 
\end{itemize}
Mathematically, an unrestricted range is equivalent to the voting function being surjective or onto. 
Having an unrestricted range can be seen as a form of the non-imposition criterion for single-winner systems. 
Note that neutrality implies an unrestricted range, so having an unrestricted range can be thought of as a relaxation of neutrality. 
Yet Gibbard and Satterthwaite have independently shown that even under such a relaxed version of neutrality there is no strategy-proof electoral system other than dictatorship~\cite{gibbard1973manipulation,satterthwaite1975strategy}. 
\begin{theorem}[Gibbard--Satterthwaite theorem]
There is no unrestricted range electoral system for an election with more than two candidates that is strategy-proof and is not dictatorial. 
\end{theorem}

The results of Gibbard and Satterthwaite further demonstrate a one-to-one correspondence between strategy-proof systems and systems satisfying Arrow's criteria. 
Duggan and Schwartz have proved a generalized version of the theorem not restricted to single-winner systems~\cite{duggan2000strategic}. 

In light of such impossibility results, and with completely strategy-proof systems out of the question, electoral systems may be examined based on the specific manipulation strategies to which they are prone. 
The choice of a system can then be made based on the occurring probability and severity of such possible manipulation strategies in the contextual circumstances of a specific election.

\subsubsection{Systems with Respect to Criteria}
\label{sec:compsyscrit}

\newcommand{\ohyes}{\ding{51}}
\newcommand{\ohno}{\ding{55}}

\begin{table}
  \centering 
  \caption{Selected electoral system and criteria they satisfy}
  \label{tbl:syscrit}
  \fbox{ 
  \begin{tabular}{lccccccc} 
    System 		& Equ.   & Neu.   & Maj.   & Con.   & Mon.   & Par.   & IIA    \\ 
    \hline 
    FPTP 		& \ohyes & \ohyes & \ohyes & \ohno  & \ohyes & \ohyes & \ohno  \\ 
    Approval 	& \ohyes & \ohyes & \ohno  & \ohno  & \ohyes & \ohyes & \ohyes \\ 
    TRS 		& \ohyes & \ohyes & \ohyes & \ohno  & \ohno  & \ohyes & \ohno  \\ 
    Contingent 	& \ohyes & \ohyes & \ohyes & \ohno  & \ohno  & \ohyes & \ohno  \\ 
    Exhaustive 	& \ohyes & \ohyes & \ohyes & \ohno  & \ohno  & \ohyes & \ohno  \\ 
    IRV 		& \ohyes & \ohyes & \ohyes & \ohno  & \ohno  & \ohyes & \ohno  \\ 
    Borda 		& \ohyes & \ohyes & \ohno  & \ohno  & \ohyes & \ohyes & \ohno  \\ 
    Cumulative 	& \ohyes & \ohyes & \ohno  & \ohno  & \ohyes & \ohyes & \ohno  \\ 
    Schulze 	& \ohyes & \ohyes & \ohyes & \ohyes & \ohyes & \ohyes & \ohno  \\ 
  \end{tabular}
  }
\end{table}

Table~\ref{tbl:syscrit} lists selected electoral systems and criteria they do and do not satisfy. 
A tick (\ohyes) indicates that the system on that row always satisfies the criterion on that column, whereas a cross (\ohno) indicates that the system does not necessarily satisfy the criterion. 
The criteria discussed in this chapter and presented in the table are a selective set of those discussed in social choice theory. 

Note that, assuming that voters do not change their minds between multiple rounds of an election, the TRS and contingent votes can be thought of as the same system in theory, and hence the two systems have the same properties in Table~\ref{tbl:syscrit}. 
The same statement is also true about the exhaustive vote and IRV systems. 

In some cases, it is easy to see why a system satisfies a specific criterion; e.g., a candidate that achieves a majority obviously achieves a plurality as well, and hence FPTP satisfies the majority criterion. 
In other cases, the reason for a tick or a cross might be less obvious. 
We leave the task of justifying the ticks to the reader, but give some counter-examples to explain some of the crosses in the following. 
Figure~\ref{fig:voex} contains the counter-examples we are going to use to this end. 
Each counter-example is a profile of an election which specifies the number of voters that have a specific candidate preference. 
For instance, the profile indicated as ``Election~1'' basically says 4 voters have the preference $A \succ B \succ C$, 2 the preference $B \succ C \succ A$, and 3 the preference $C \succ B \succ A$. 


Consider Election~1 in Figure~\ref{fig:voex}. 
A FPTP election would record 4 votes for A, 2 for B, and 3 for C, and hence the FPTP winner would be A. 
However, in one-on-one elections, B would beat both A and C, 5--4 and 6--3, respectively, and hence B is the Condorcet winner. 
In fact even C beats A 5--4 in a head-to-head election, which means FPTP might even elect a \emph{Condorcet loser}, i.e., a candidate that loses against all other candidates in head-to-head elections. 
Also note that if the third group change their preference from $C \succ B \succ A$ to $B \succ C \succ A$, the winner of FPTP will change to B, despite the fact that the voters who have changed their mind still rank A the same way with respect to B and C, i.e., they still think $C \succ A$ and $B \succ A$. 
Thus FPTP does not satisfy the Condorcet and IIA criteria. 

Consider Election~2 in Figure~\ref{fig:voex} from~\cite{pacuit2012voting}. 
With either two-round system (TRS) or instant run-off voting (IRV), C is eliminated in the first round, and in the second round between A and B, C's votes go to A and hence A wins the TRS or IRV elections. 
Now consider the case where A is able to gain the support of the last group of 2 voters and change their preference to $A \succ B \succ C$. 
In that case, B gets eliminated in the first round, and in the second round C beats A 9--8. 
Thus, A loses the second election despite gaining votes. 
This shows that TRS and IRV (and hence the contingent and exhaustive vote systems) are not monotone. 

If any of the four systems above, i.e., TRS, IRV, contingent, or exhaustive, is used to elect the winner in Election~1 in Figure~\ref{fig:voex}, the Condorcet winner B will be eliminated in the first round and C will be the eventual winner. 
Hence, these systems do not necessarily elect the Condorcet winner. 

Consider Election~3 in Figure~\ref{fig:voex} from Condorcet~\cite{condorcet1785essai}. 
It is not hard to see that A is the Condorcet winner but using the standard Borda count, i.e., allocating 2, 1, and 0 points for the 1st, 2nd, and 3rd preferences, respectively, elects B as the winner. 
In fact, even in a generalized Borda count where $p_i$ points are allocated for the $i$-th preference, A receives $31 p_1 + 39 p_2 + 11 p_3$ points and B $39 p_1 + 31 p_2 + 11 p_3$ points. 
Since $p_1$ needs to be greater than $p_2$ for the system to make sense, this example shows that no generalized Borda count can guarantee electing the Condorcet winner. 

\begin{figure}
\quad
\begin{minipage}{.24\linewidth}
\centering 
Election 1 \\ 
\fbox{
\begin{tabular}{lll}
  4 & 2 & 3 \\ \hline 
  A & B & C \\ 
  B & C & B \\ 
  C & A & A \\ 
\end{tabular} 
}
\end{minipage}%
\begin{minipage}{.27\linewidth}
\centering 
Election 2 \\ 
\fbox{
\begin{tabular}{llll}
  6 & 5 & 4 & 2 \\ \hline 
  A & C & B & B \\ 
  B & A & C & A \\ 
  C & B & A & C \\ 
\end{tabular} 
}
\end{minipage}%
\begin{minipage}{.42\linewidth}
\centering 
Election 3 \\ 
\fbox{
\begin{tabular}{llllll}
  30 & 1 & 29 & 10 & 10 & 1 \\ \hline 
  A & A & B & B & C & C \\ 
  B & C & A & C & A & B \\ 
  C & B & C & A & B & A \\ 
\end{tabular} 
}
\end{minipage}%
\caption{Counter-examples of election profiles}
\label{fig:voex}
\end{figure}

Approval voting is a bit trickier in that the outcome of the election not only depends on voter preferences, but also on the number of candidates each voter approves. 
This means, unlike some other systems such as FPTP, TRS, and Borda, in an approval voting election for each election profile there might be multiple possible outcomes based on voters' behaviour. 
For instance, in Election~1 in Figure~\ref{fig:voex}, if all voters only approve their top candidate, A would win the election, whereas if all voters approve their top two candidates, B would win, and at the same time, if the first and third groups of voters approve one candidate and the second group approves two, then C would win. 
A similar situation may happen even if a candidate has a majority. 
Thus, approval voting without any restriction on how many candidates may be approved by voters does not satisfy the majority and Condorcet criteria. 

All counter-examples used for FPTP and Borda may be also used for cumulative voting since both FPTP and Borda can be seen as instances of cumulative voting.

\subsection{A Practitioner's Point of View}
\label{sec:syscomppr}

In practice, electoral systems are usually broadly categorized as \emph{majoritarian}, \emph{proportional}, and \emph{mixed} systems. 
Majoritarian systems are based on the general principle that a single candidate with a plurality of votes is elected to represent and pursue the demands of a specific (usually geographic) constituency. 
FPTP, TRS, IRV, and other similar systems hence fall in this category. 
Proportional systems on the other hand, are based on the general principle that the elected body of candidates proportionally reflects the diverse range of views in a heterogeneous society. 
This category includes multiple list voting systems and STV, although STV is sometimes referred to as semi-proportional. 
Mixed systems aim to attain the best of both worlds by incorporating elements from the above two types of systems. 
MMP and two-tier systems are examples of mixed systems. 
This categorization is a general guide and some systems, most notably SNTV, do not seem to fit in any of the categories. 

The underlying principles of the majoritarian and proportional systems correspond to two different conceptions of ``representation'': \emph{principal--agent} and \emph{microcosm}, as put forth by McLean~\cite{mclean1991forms}. 
The principal--agent conception defines representation as an agent acting on behalf of a principal, whereas the microcosmic conception defines representation as statistically typifying the group being represented. 
McLean argues that the two conceptions are each entirely reasonable but inconsistent with each other. 

Rae distinguished three main components of an electoral system: \emph{district magnitude}, \emph{electoral formula}, and \emph{ballot structure}~\cite{rae1967political}. 
District magnitude refers to the number of candidates elected in each electoral district; 
electoral formula is the algorithm used to calculate the winner(s); and 
ballot structure refers to the information collected from the voter on a ballot. 
Rae further argues that classification of electoral systems often deals with only one component, namely the electoral formula, and leaves the other two out, whereas district magnitude and ballot structure have significant effects on how an electoral system performs. 
Based on district magnitude, systems can be classified into single-member and multi-member district systems. 
Different ballot structures on the other hand lead to categorizing systems based on three aspects: 
first, the number of votes allowed: either one, more than one but less than the number of seats or equal to the number of candidates or seats; 
second, the type of information the voter is asked to provide: either nominal, ordinal, or cardinal, 
and third, for whom the voter votes: either for individuals or for groups of individuals (e.g., parties)~\cite{blais1988classification}. 
Systems using nominal ballots (i.e., voting for one option) include FPTP, TRS, and closed-list PR; 
systems using ordinal ballots (i.e., ranking the options) include IRV, STV, and Borda count; and 
systems using cardinal ballots (i.e., rating the options) include approval and range voting. 

Majoritarian systems are praised for their ability to produce a clear tie between an elected candidate and a constituency, which in turn implies a clear responsibility and accountability of the elected candidate towards the constituency. 
Besides, most majoritarian systems (with, e.g., IRV being an exception) are simple to understand and do not require complex mathematics to calculate the results, and hence they are considered to encourage transparency. 
However, such systems tend to favour large parties and do not usually produce results that reflect the shares of votes received by different parties. 
Thus, minority groups and smaller parties may not be able to win any seat and are encouraged to integrate into the larger parties. 
In some contexts, e.g., when there are two dominant parties, this can be seen as a positive feature since it produces a clear winner and hence a strong and stable government as well as a strong opposition and government alternative. 

Proportional systems on the other hand emphasize accurately representing the make-up of diverse electorates. 
The greater the number of candidates to be elected from an electoral district, the more proportional the results tend to be. 
Such systems should result in governing coalitions that represent a wide range of views in the political scene, although in some contexts, negotiations to build a coalition may take a long time. 
Proportional systems tend to facilitate fragmentation of the party system. 
Besides, since multi-member districts are required to guarantee any degree of proportionality, proportional systems usually lack the clear link between a specific candidate and the constituency. 
In contrast with proportionality, the greater the number of candidates to be elected from an electoral district, the weaker such links tend to be. 

While the principal--agent and microcosmic conceptions describe an elected body's collective role in representing the electors, an elected candidate's individual representative role may be defined as that of either a \emph{delegate} or a \emph{trustee}. 
A delegate in this characterization is expected to listen to and reflect the views of the electors, whereas a trustee is thought to be entrusted by the electors to use his or her own judgement and decide on behalf of the electors. 
Farrell argues that in ``party-based'' electoral systems there is a greater tendency for elected representatives to act as trustees, whereas comparatively in ``candidate-based'' systems there is more incentive for elected representatives to act as delegates~\cite{farrell2011electoral}. 

Majoritarian systems are considered more susceptible to strategic voting compared to proportional systems. 
In a FPTP system for example, a voter might vote for a candidate that they do not prefer but think has a better chance to win. 
Proportional systems, on the other hand, are considered to encourage voters to declare their actual preferences. 

Majoritarian systems, especially those using single-member districts, are prone to district boundary irregularities, known as \emph{malapportionment} and \emph{gerrymandering}, that might arise as a result of the process of district delimitation~\cite[pp.~202--205]{farrell2011electoral}. 
Malapportionment refers to the situations in which there are imbalances between the populations of different electoral districts that favour one party over others. 
Gerrymandering refers to the practice of (re)drawing electoral boundaries in shapes that are expected to disproportionately boost the number of seats won by a specific party. 
Some proportional systems, specially those using smaller multi-member districts, are susceptible to such irregularities as well. 
Generally speaking, the greater the number of candidates to be elected in districts, the less they have the potential to suffer from malapportionment and gerrymandering~\cite[Ch.~10]{hodge2005mathematics}. 
These issues however may be resolved by putting a neutral body in charge of district delimitation. 

A widely accepted characterization is that of Duverger who argues that the single-ballot plurality systems favour party dualism, whereas two-round majority systems and proportional systems favour multipartism~\cite{duverger1963political}. 
He further argues that majoritarian systems may encourage ``personality parties'', i.e., those based on a leader's popularity, and geographic minority parties, whereas proportional systems generally encourage ``permanent minority parties'', such as ethnic or religious ones, but discourage ``personality parties''. 
The effects of mixed systems are less understood as these systems only relatively recently have been adopted by a considerable number of countries. 

Proportional systems tend to be more accommodating in adjusting representation towards historically under-represented groups and minorities. 
In established democracies, systems based on multi-member districts have shown a strong increase in women's representation, whereas this trend is much weaker in systems based on single-member districts~\cite{matland2005enhancing}. 

Votes that do not count towards the election of any candidate are usually referred to as \emph{wasted} votes. 
Systems such as FPTP tend to leave a larger number of wasted votes, whereas proportional systems with low thresholds, IRV, and STV aim to reduce the number of wasted votes. 
A related issue is \emph{vote splitting}, and it happens when similar candidates compete in an election and their potential supporters' votes tend to be split between them, which possibly allows a candidate representing a less popular overall viewpoint to win. 
FPTP particularly suffers from this issue, whereas TRS is considered less susceptible, and proportional systems with low thresholds, IRV, and STV are considered relatively immune to vote splitting. 

The two-round system is unique among the discussed systems in that it possibly requires the electoral administration to run a second election in a short period, hence significantly increasing the election cost. 
On the other hand, this unique property enables voters to change their minds from the first round to the second and accelerate consensus building between parties to coalesce behind the candidates in the second round. 

Among the multiple highest average (HA) seat allocation methods for list electoral systems, the Danish method is considered to comparatively favour smaller parties; 
the Sainte-Lagu\"{e} method is considered neutral; 
the modified Sainte-Lagu\"{e} and Imperiali methods are considered to favour larger parties; and 
the d'Hondt method is considered to favour larger parties the most. 
Among the largest remainder (LR) methods, smaller quotas are more favourable to larger parties. 
Considering all proportional systems, it has been shown that they can be generally ordered from the most to the least favourable to the larger parties as follows~\cite{gallagher1992comparing}: LR using Imperiali quota, d'Hondt, STV, LR using Droop quota, modified Sainte-Lagu\"{e}, LR using Hare quota and Sainte-Lagu\"{e}, and finally the Danish method. 

Mixed systems tend to produce election results that, in terms of proportionality, fall between majoritarian and proportional systems. 
However, some criticize such systems for effectively creating two classes of elected candidates with different mandates and hence undermining the cohesiveness of the elected body of representatives. 

Among the systems that do not fall in the three categories mentioned above, SNTV is considered to be easy to understand, to accommodate the representation of minority parties better compared to majoritarian systems, and to fragment the party system less compared to proportional systems. 
However, SNTV tends to result in many wasted votes, and parties need to consider complex strategic decisions as to how many candidates to put forth as the system suffers from issues similar to vote splitting.

\section*{Acknowledgement}
The author is supported by the ERC Starting Grant No.~306994.





{\small 
\bibliographystyle{plain} 
\bibliography{bookref,bookrefextra}
}

\end{document}